\documentclass[a4paper,10pt]{article} 
\input epsf.tex

\newcommand{\be}{\begin{equation}}
\newcommand{\ee}{\end{equation}}
\newcommand{\ba}{\begin{eqnarray}}
\newcommand{\ea}{\end{eqnarray}}
\newcommand{\ban}{\begin{eqnarray*}}
\newcommand{\ean}{\end{eqnarray*}}

\newcommand{\braket}[2]{\mbox{$ \langle #1 | #2 \rangle $}}
\newcommand{\sandwich}[3]{\mbox{$ \langle #1 | #2 | #3 \rangle $}}
\newcommand{\ket}[1]{\mbox{$ | #1 \rangle $}}

\newcommand{\com}[2]{\left[ #1\,,\,#2 \right]}

\newcommand{\si}{\sigma}
\newcommand{\demi}{\frac{1}{2}}
\newcommand{\compl}{\begin{picture}(8,8)\put(0,0){C}\put(3,0.3){\line(0,1){7}}\end{picture}}

\newcommand{\one}{\leavevmode\hbox{\small1\normalsize\kern-.33em1}}

\newcommand{\moy}[1]{\langle #1 \rangle}

\begin{document}

\title{\Large \sc{Spectral decomposition of Bell's operators for qubits}}
{\normalsize{\author{ Valerio Scarani\thanks{corresponding
author. Tel. +41 22 7026883; fax +41 22 7810980; e-mail:
valerio.scarani@physics.unige.ch}, Nicolas Gisin\\
Group of Applied Physics, University of Geneva\\
20, rue de l'Ecole-de-M\'edecine, CH-1211 Geneva, Switzerland\\
 }}}
\maketitle

\begin{abstract}
The spectral decomposition is given for the N-qubit Bell
operators with two observables per qubit. It is found that the
eigenstates (when non-degenerate) are N-qubit GHZ states even for
those operators that do not allow the maximal violation of the
corresponding inequality. We present two applications of this
analysis. In particular, we discuss the existence of pure
entangled states that do not violate any Mermin-Klyshko
inequality for $N\geq 3$.
\end{abstract}

\section{Introduction}

There is a huge literature about the topic of Bell's
inequalities. For the sake of this introduction, we briefly recall
some well-known ideas, using the Clauser-Horne-Shimony-Holt (CHSH)
inequality \cite{chshpaper}. For each choice of four numbers
$a_1,a'_1,a_2,a_2'\in\{-1,+1\}$, the quantity
$S=\demi(a_2+a_2')a_1+\demi(a_2-a'_2)a_1'$ can take only the
values +1 or -1. Therefore, if the four numbers are considered as
realization of random variables, the expectation of $S$ will
certainly depend on the distribution of the variables but must
satisfy $|E(S)|\leq 1$. These are "trivial mathematics". But if
one turns to quantum mechanics (QM), then this inequality can be
violated. Indeed, consider that $a$ is $+1$ if, as a result of a
measurement, a spin is found along the direction $+\mathbf{a}$,
and $a$ is $-1$ if the spin is found along the direction
$-\mathbf{a}$. This is achieved by replacing $a$ by the operator
$\mathbf{a}\cdot\mathbf{\sigma}\equiv\sigma_a$. Using this
prescription for $a_1$, $a_1'$, $a_2$ and $a_2'$ we find that $S$
is the expectation value of the "Bell operator" \cite{note0} \be
B_2(\mathbf{a_1},\mathbf{a_1'}, \mathbf{a_2},\mathbf{a_2'})\,=\,
\demi\left(\sigma_{a_2}+\si_{a'_2}\right)\otimes\si_{a_1} +
\demi\left(\sigma_{a_2}-\si_{a'_2}\right)\otimes\si_{a_1'}
\label{chsh}\ee where $\mathbf{a_1},\mathbf{a_1'},
\mathbf{a_2},\mathbf{a_2'}$ are unit vectors that will be referred
to as the "parameters" of the Bell operator. It is well-known
that for some choices of the parameters the highest eigenvalue of
$B_2$ can be higher than 1: there exist some states $\ket{\Psi}$
that violate the inequality $|\moy{B_2}_{\Psi}|=|E(S)|\leq 1$.
"Trivial mathematics" fail whenever
$\big[\si_{a_k},\si_{a_k'}\big]\neq 0$, that is whenever
$\mathbf{a_k}\neq\pm\mathbf{a_k'}$: in this case the "random
variables" $a_k$ and $a_k'$ cannot have simultaneously a precise
value.

The first Bell's inequalities were derived for two two-level
systems (hereafter referred to as "qubits"). Generalizations of
the Bell's inequalities have been proposed along the following
lines: (i) bipartite inequalities for two n-level quantum systems
\cite{gisinperes}; (ii) bipartite inequalities using more than
two parameters per system $a_1$, $a_1'$, $a_1''$,... \cite{brau};
(iii) multipartite inequalities, that is, inequalities involving
more than two quantum systems \cite{mermin,belin,helle}.

In this paper, we consider inequalities involving an arbitrary
number of qubits using two observables per qubit (the observables
are obviously dichotomic). This family of inequalities has been
studied in great detail independently by Werner and Wolf \cite{ww}
and by Zukowski and Brukner \cite{zb}. Our present contribution
consists in exhibiting explicitly the spectral decomposition of
the Bell's operators (section 2). Section 3 presents applications
of this result.

\section{Spectral decomposition of Bell operators}

\subsection{Inequalities for two observables}

Consider a quantum system composed of $n$ qubits, that is a
system described by the Hilbert space
$\big(\compl^2\big)^{\otimes n}$. For each qubit $k$, we define
two observables $A_{k}(0)=\si_{a_k}$ and $A_{k}(1) = \si_{a_k'}$,
with $\mathbf{a_k}$ and $\mathbf{a_k'}$ two vectors on the unit
sphere. The set $\{\mathbf{a_1},\mathbf{a_1'},...
\mathbf{a_n},\mathbf{a_n'}\}$ of the $2n$ unit vectors is written
$\underline{\mathbf{a}}$. Up to normalization, any n-qubit Bell
inequality\footnote{From now onwards, expressions like "all
inequalities" mean "all inequalities involving two observables
per qubit".} can be written as $\moy{{\cal{B}}_n
(\underline{\mathbf{a}})} \leq 1$ for a given Bell operator
${\cal{B}}_n$. The form of the Bell operator is a polynomial \ba
{\cal{B}}_n(\underline{\mathbf{a}}) &=&
\sum_{s\in\{0,1\}^n}\,\beta(s)\,\bigotimes_{k=1}^n A_k(s_k)\,.
\label{polynomial}\ea The coefficients $\beta(s)$ are rather
arbitrary, provided that $\moy{{\cal{B}}_n} \leq 1$ is satisfied
for all product states.

Of course, not every polynomial of the form (\ref{polynomial})
defines a good inequality; in the worst cases, e.g. when the
polynomial is simply $\si_{a_1}\otimes...\otimes\si_{a_n}$, one
will find $\moy{{\cal{B}}_n} \leq 1$ for all states. The complete
classification of all inequalities is due to Werner and Wolf
\cite{ww}. A special role is played by the Mermin-Klyshko (MK)
inequalities \cite{mermin,belin,helle}, whose corresponding Bell
operator is defined recursively as \ba
B_n(\underline{\mathbf{a}})\,\equiv\,B_n&=&\demi(\si_{a_n}+\si_{a_n'})\otimes
B_{n-1}\,+\, \demi(\si_{a_n}-\si_{a_n'})\otimes B_{n-1}'
\label{recurrence}\ea where $B_n'$ is obtained from $B_n$ by
exchanging all the $\mathbf{a}_k$ and $\mathbf{a}_k'$. In
particular, $B_2$ is given by the CHSH inequality; $B_3$ is the
operator that corresponds to the so-called Mermin's inequality
\cite{mermin}. For MK inequalities, the violation allowed by QM is
$\moy{B_n}=2^{(n-1)/2}$; no other inequality with two observables
per qubit can reach such a violation \cite{ww}. More results on
these operators are given in Appendices A and B.

\subsection{Spectral decomposition: statement of the theorem}

We want to characterize the eigenvectors and eigenvalues of the
Bell operator ${\cal{B}}_n$ defined in (\ref{polynomial}), for a
given set of $2n$ unit vectors $\underline{\mathbf{a}}$. For this
purpose, we can suppose without any loss of generality that all
the unit vectors in $\underline{\mathbf{a}}$ lie in the $(x,y)$
plane; physically, this amounts to say that the axes $x,y,z$ can
be defined independently for each qubit. We will show the
following
\newtheorem{propeigen}{Theorem}
\begin{propeigen}
Let ${\cal{B}}_n$ given by (\ref{polynomial}), with
$\mathbf{a_k}=\cos\alpha_k \mathbf{e_x}+\cos\alpha_k
\mathbf{e_y}$ and $\mathbf{a_k'}=\cos\alpha'_k
\mathbf{e_x}+\cos\alpha'_k \mathbf{e_y}$ for all $k=1,...,n$. Let
$\ket{0}$, resp. $\ket{1}$, be the eigenvector of $\si_z$ for the
eigenvalue $+1$, resp. $-1$. Finally, let
$\Omega=(\omega_1,...,\omega_n)\,\in\{0,1\}^n$ be a configuration
of $n$ zeros or ones, and
$\bar{\Omega}=(\bar{\omega}_1,...,\bar{\omega}_n)$ with
$\bar{\omega}_k=1-\omega_k$ the complementary configuration. Then:
\begin{enumerate}
\item The $2^n$ n-qubit GHZ states, labeled by the configurations
$\Omega$, defined by: \ba
\ket{\Psi_{\Omega}(\theta_{\Omega})}\,=\, \frac{1}{\sqrt{2}}\,
\left(e^{i\theta_{\Omega}}\ket{\Omega}\,+\,
\ket{\bar{\Omega}}\right)\label{psitheta}\ea form a basis of
eigenvectors of ${\cal{B}}_n$ for some $\theta_{\Omega}
=\theta_{\Omega}(\underline{\mathbf{a}})$.
\item The parameter $\theta_{\Omega}
=\theta_{\Omega}(\underline{\mathbf{a}})$ and the eigenvalue
$\lambda_{\Omega}= \lambda_{\Omega}(\underline{\mathbf{a}})$ are
calculated from a complex number
$f_{\Omega}(\underline{\mathbf{a}})$: \be
\begin{array}{llll}\mbox{if }\arg f_{\Omega}\in[0,\pi[& \mbox{then:} &\theta_{\Omega}=-\arg
f_{\Omega}-\pi\,,&\lambda_{\Omega}=-|f_{\Omega}|\,;\\
\mbox{if }\arg f_{\Omega}\in[\pi,2\pi[& \mbox{then:}
&\theta_{\Omega}=-\arg
f_{\Omega}\,,&\lambda_{\Omega}=|f_{\Omega}|\,.\end{array}\label{solution}\ee
The complex number $f_{\Omega}(\underline{\mathbf{a}})$ is
obtained as follows: take ${\cal{B}}_n$, and for all $k=1,...,n$
replace the operator $\si_{a_k}$ by the complex number
$e^{i\alpha_k}$ if $\omega_k=0$ in $\Omega$, or by the complex
number $e^{-i\alpha_k}$ if $\omega_k=1$ in $\Omega$; and the
analog replacement for $\si_{a_k'}$.
\end{enumerate}
\label{propeigen}
\end{propeigen}
About statement 1: It was noticed in \cite{ww} (V, D) that GHZ
states are the states that maximally violate any inequality. This
is an immediate corollary of statement 1, since for any matrix
$M$ it holds that $\max_v\sandwich{v}{M}{v}/\braket{v}{v}$ is the
maximal eigenvalue, obtained if and only if $v$ is the
eigenvector associated to that eigenvalue.

About statement 2: The definition of
$f_{\Omega}(\underline{\mathbf{a}})$ may seem cumbersome, but a
single example will clarify it. Take the CHSH operator is given by
(\ref{chsh}). Then $f_{00}=\demi(e^{i\alpha_2}+e^{i\alpha_2'})
\,e^{i\alpha_1}\,+\,\demi(e^{i\alpha_2}-e^{i\alpha_2'})\,e^{i\alpha_1'}$;
$f_{01}$ is obtained by replacing $e^{i\alpha_2}$ and
$e^{i\alpha_2'}$ by their conjugates; $f_{10}=f_{01}^{*}$ and
$f_{11}=f_{00}^{*}$, with $z^{*}$ the conjugate of $z$.

The proof of the theorem is given in two steps. In the first
step, we take advantage of a remarkable symmetry of the Bell
operators to guess the basis (\ref{psitheta}); in the second step,
direct calculation allows to get the announced explicit results
(\ref{solution}).

\subsection{First step}

For a given rotation matrix $R\in SO(3)$, it is well-known that
one can find $U\in SU(2)$ such that
$U\,\mathbf{a}\cdot\si\,U^{-1}\,=\,(R\mathbf{a})\cdot\si$. In
particular, one can find $U$ such that $U\si_a
U^{-1}=\si_{-a}=-\si_a$. Since we are considering that all the
parameters $\underline{\mathbf{a}}$ lie in the plane $(x,y)$, the
rotation that brings $\mathbf{a}$ on $-\mathbf{a}$ is a rotation
by $\pi$ around the $z$-axis, so that the corresponding unitary
operation is $U\simeq\si_z$ (equality up to an arbitrary phase).
We introduce the notation $U[k]=\one\otimes...\otimes
\si_z\otimes...\otimes\one$, where the rotation is applied on the
$k$-th qubit. Note that $\com{U[k]}{U[l]}=0$ for all $k$ and $l$.
Since ${\cal{B}}_n$ is a sum of terms like
$\si_{a_1}\si_{a_2'}...\si_{a_n}$, we have manifestly: \ba
U[k]\,{\cal{B}}_n\,U[k]^{-1}\,=\,-{\cal{B}}_n&&\forall k\in\{1,...,n\}\label{unu}\\
U[k]U[l]\,{\cal{B}}_n\,U[k]^{-1}U[l]^{-1}\,=\,{\cal{B}}_n&&\forall\,k,l\in\{1,...,n\}\,.\label{deuxu}
\ea These conditions depend critically on the assumption that the
Bell operator is dichotomic: if we had three or more vectors for a
qubit, we could not ensure that they lie in a plane. Condition
(\ref{unu}) says that ${\cal{B}}_n$ and $-{\cal{B}}_n$ are linked
by a unitary operation, whence the following:
\newtheorem{l4}{Lemma}
\begin{l4}
If $\lambda$ is an eigenvalue of ${\cal{B}}_n$ associated to
$\ket{\psi}$, then $-\lambda$ is also an eigenvalue of
${\cal{B}}_n$. The vector $U[k]\ket{\psi}$ is eigenvector of
${\cal{B}}_n$ for the eigenvalue $-\lambda$, for all $k$.
\label{l4}
\end{l4}
The symmetries (\ref{unu}) and (\ref{deuxu}) of ${\cal{B}}_n$
suggest to look for vectors satisfying \ba
U[k]\ket{\Psi}&\perp\,\ket{\Psi}&\forall k\leq n\,;\label{condition1}\\
U[k]U[l]\ket{\Psi}&=\,e^{i\gamma_{kl}}\ket{\Psi}&\forall k,l\leq
n\label{condition2} \ea as good candidates for the eigenstates of
${\cal{B}}_n$ --- they would be the unique candidates if none of
the eigenvalues of ${\cal{B}}_n$ were degenerate, but this is
generally not the case (see Appendix B). Let $\ket{0}$ (resp.
$\ket{1}$) be the eigenstate of $\si_z$ for the eigenvalue +1
(resp. -1). We decompose the n-qubit state $\ket{\Psi}$ on the
basis of the product states of $\ket{0}$s and $\ket{1}$s:
$\ket{\Psi}\,=\, \sum_{\Omega\in\{0,1\}^n}
c_{\Omega}\ket{\Omega}$. We use condition (\ref{condition2})
first: \ba
U[k]U[l]\ket{\Psi}&=&\sum_{\Omega}(-1)^{\omega_k+\omega_l}
c_{\Omega}\ket{\Omega}\,
\stackrel{(\ref{condition2})}{=}\,e^{i\gamma_{kl}}\sum_{\Omega}
c_{\Omega}\ket{\Omega}\;\forall k,l\;\Longleftrightarrow\nonumber\\
&\Longleftrightarrow& \big[e^{i\pi(\omega_k+\omega_l)}-
e^{i\gamma_{kl}}\big]\,c_{\Omega}\,=\,0\;\;
\forall\,\Omega\in\{0,1\}^n\;\mbox{ and }\;\forall\,k,l\leq n\,.
\label{cond2explicite}\ea Now suppose $c_{\Omega}\neq 0$: this
implies, modulo $2\pi$: \ba
\gamma_{kl}\,=\,\pi(\omega_k+\omega_l)&=&
\left\{\begin{array}{lll}0&\mbox{if}&\omega_k=\omega_l\\
\pi&\mbox{if}&\omega_k\neq\omega_l
\end{array}\right.\,.\label{omega}\ea
That is, the choice of $\Omega$ for which $c_{\Omega}\neq 0$
determines completely the sequence of the $\gamma_{kl}$. Now, it
is evident from (\ref{omega}) that only $\bar{\Omega}$ gives
exactly the same sequence as $\Omega$. Thus
(\ref{cond2explicite}) means that once we have chosen $\Omega$
for which $c_{\Omega}\neq 0$, then $c_{\Omega'}=0$ for all
$\Omega'\,\neq\,\Omega,\,\bar{\Omega}$. We turn now to condition
(\ref{condition1}), that, with
$U[k]\ket{\Psi}=\sum_{\Omega}(-1)^{\omega_k}
c_{\Omega}\ket{\Omega}$, reads: \ba
\braket{\Psi}{U[k]\Psi}&=&\sum_{\Omega}(-1)^{\omega_k}
|c_{\Omega}|^2\,\stackrel{(\ref{condition1})}{=}\,0\;\forall
k\,.\label{cond1expl}\ea But we have proved just above that the
states we are interested in are such that only $c_{\Omega}$ and
$c_{\bar{\Omega}}$ can be different from zero. Thus
(\ref{cond1expl}) becomes $(-1)^{\omega_k}\,(|c_{\Omega}|^2-
|c_{\bar{\Omega}}|^2)=0$, that is $|c_{\Omega}|=
|c_{\bar{\Omega}}|$. We have then proved that a n-qubit state
satisfies both (\ref{condition1}) and (\ref{condition2}) if and
only if it is of the form (\ref{psitheta}) for a given
$\Omega\in\{0,1\}^n$. Thus we have $2^n$ states, each labeled by
one configuration $\Omega$. The orthogonality requirement
$\braket{\Psi_{\Omega}(\theta_{\Omega})}{\Psi_{\Omega'}(\theta_{\Omega'})}
=\delta_{\Omega,\Omega'}$ is trivial but for
$\Omega'=\bar{\Omega}$: in this case, we must require
$\theta_{\bar{\Omega}}=\pi- \theta_{\Omega}$. This concludes the
first step of the proof. Just two remarks before turning to the
second step:

{\em Remark 1:} The state built on $\bar{\Omega}$ is entirely
determined by the state built on $\Omega$ through \be
\ket{\Psi_{\bar{\Omega}} (\theta_{\bar{\Omega}})}\,\simeq\,
\ket{\Psi_{\Omega}(\theta_{\Omega}+\pi)}\,\simeq\,
U[k]\ket{\Psi_{\Omega}(\theta_{\Omega})}\,. \label{3charact}\ee
The first equality follows from the requirement
$\theta_{\bar{\Omega}}=\pi- \theta_{\Omega}$ by extracting
$\theta_{\bar{\Omega}}$ as a global phase. As for the second
equality: \ban U[k]\ket{\Psi_{\Omega}(\theta_{\Omega})}&=&
\frac{1}{\sqrt{2}}\,
\left(e^{i\theta_{\Omega}}(-1)^{\omega_k}\ket{\Omega}\,+\,
(-1)^{\bar{\omega}_k}\ket{\bar{\Omega}}\right)\,=\\
&=&\frac{e^{i\theta_{\Omega}}(-1)^{\omega_k}}{\sqrt{2}}\,
\left(\ket{\Omega}\,-\,
e^{-i\theta_{\Omega}}\ket{\bar{\Omega}}\right)\,\simeq\,
\frac{1}{\sqrt{2}}\,
\left(e^{i(\pi-\theta_{\Omega})}\ket{\bar{\Omega}} \,+\,
\ket{\Omega}\right)\,.\ean Thus,
another recipe to build a basis of states of the form (\ref{psitheta}) is the following: (i)
for all $\Omega$ such as (say) $\omega_1=0$, choose
$\theta_{\Omega}$ and build the state
$\ket{\Psi(\Omega,\theta_{\Omega})}$;
(ii) apply $U[k]$ to each of these states (or change
$\theta_{\Omega}$ to $\theta_{\Omega}+\pi$) to complete the set.

{\em Remark 2:} The two states built from $\Omega_0=(0,...,0)$,
namely $\frac{1}{\sqrt{2}}\,
\left(e^{i\theta}\ket{0...0}\,\pm\,\ket{1...1}\right)$, are the
only ones for which the $\gamma_{kl}$ are independent of $k$ and
$l$ --- and are actually 0, with our choice of phases
$U[k]=\si_z[k]$.

\subsection{Second step}

We must show that the states of the form (\ref{psitheta}) form a
basis of eigenstates of ${\cal{B}}_n(\underline{\mathbf{a}})$,
that is \be {\cal{B}}_n
\,\ket{\Psi_{\Omega}(\theta_{\Omega})}\,=\,\lambda_{\Omega}
\,\ket{\Psi_{\Omega}(\theta_{\Omega})} \label{eigen1}\ee with
$\theta_{\Omega}=\theta_{\Omega}(\underline{\mathbf{a}})$ and
$\lambda_{\Omega}= \lambda_{\Omega}(\underline{\mathbf{a}})$.
Actually, we have to solve (\ref{eigen1}) only for the state
built on $\Omega_0=(0,...,0)$\ban
\ket{\Psi_{\Omega_0}(\theta_{\Omega_0})} \,\equiv\,
\ket{\Psi(\theta)}&=& \frac{1}{\sqrt{2}}\,
\left(e^{i\theta}\ket{0...0}\,+\,\ket{1...1}\right)\,. \ean This
is so, because one can exchange $\ket{0}$ and $\ket{1}$ by
applying a unitary operation, here $\sigma_x$. Therefore by
application of $\si_x$ to the suitable qubits we can always
transform any $\ket{\Psi_{\Omega}}$ into $\ket{\Psi_{\Omega_0}}$.
Once we have the results for $\Omega_0$, the results for $\Omega$
follow by taking the qubits $k$ for which $\omega_k=1$ in
$\Omega$, and replacing $\si_{a_k}$ by $\si_x\si_{a_k}\si_x$, that
is, replacing $(a_k)_y=\sin\alpha_k$ by $-\sin\alpha_k$, and the
same for $\si_{a_k'}$. So the eigenvalue problem is reduced to
the problem of finding $\lambda$ and $\theta$ satisfying \ba
{\cal{B}}_n\big(e^{i\theta}\ket{0...0}+\ket{1...1}\big)&=&\lambda\,
\big(e^{i\theta}\ket{0...0}+\ket{1...1}\big)\,.
\label{eigenproblem}\ea Consider now one of the terms in
(\ref{polynomial}), say $\si_{a_1}\otimes...\otimes\si_{a_n}$: a
standard calculation gives
\ban\si_{a_1}\otimes...\otimes\si_{a_n}\big(e^{i\theta}
\ket{0...0}+\ket{1...1}\big) &=& e^{i\theta} \big(\prod_k
e^{i\alpha_k}\big) \ket{1...1}+\big(\prod_k e^{-i\alpha_k}\big)
\ket{0...0}\,.\ean Consequently the eigenvalue problem
(\ref{eigenproblem}) gives \ba
{\cal{B}}_n\,\ket{\Psi(\theta)}\,=\,\frac{1}{\sqrt{2}}\,\left(e^{i\theta}f
\,\ket{1...1}\,+\,f^{*}\ket{0...0} \right)\,
=\,\lambda\ket{\Psi(\theta)} &\Longleftrightarrow&
e^{i\theta}f\,=\,\lambda \label{solimplicite}\ea with
$f(\underline{\mathbf{a}})\,=\,\big(\sum_s \beta(s) \prod_k
e^{i\alpha_k(s_k)}\big)$ --- for ease of notation, we wrote
$\alpha_k(0)$ for $\alpha_k$ and $\alpha_k(1)$ for $\alpha_k'$.
The proof of Theorem \ref{propeigen} is virtually concluded. The
solution (\ref{solution}) follows by settling a matter of
convention, since condition (\ref{solimplicite}) can be written as
$|f|e^{i(\theta+\arg(f))}=\lambda$ or as
$|f|e^{i(\theta+\arg(f)+\pi)}=-\lambda$; in other words, a
convention on $\theta$ fixes the sign of $\lambda$ --- this is
nothing but the manifestation of (\ref{3charact}). We choose as a
convention that $\theta_{\Omega}\in [0,\pi[$ for all $\Omega$;
this convention is consistent with $\theta_{\bar{\Omega}}=
\pi-\theta_{\Omega}$.

\section{Applications and perspectives}

\subsection{On some non-maximally entangled states}

In this section we study the violation of MK inequalities for a
family of N-qubit states that clearly exhibit N-qubit
entanglement. These states are \ba \ket{\psi_N(\phi)}
&=&\cos\phi\,\ket{0^N}\,+\, \sin\phi\,\ket{1^N}\,, \label{psinphi}
\ea where we adopt the notation $\ket{0^N}=\ket{0...0}$; by
convention, we choose $\cos\phi\geq \sin\phi\geq 0$ i.e.
$\phi\in[0,\frac{\pi}{4}]$.

In the case $N=2$, using Schmidt's decomposition {\em every} pure
state can be written in the form
$\ket{\psi(\phi)}=\cos\phi\ket{00}+ \sin\phi\ket{11}$. It is
well-known that the CHSH inequality is violated by all pure
entangled states \cite{gisin}; in fact, using Horodeckis' theorem
\cite{horo} one can calculate explicitly that
$S_2=\max_{\underline{a}} \moy{B_2(\underline{\mathbf{a}})}_{\psi
(\phi)} =\sqrt{1+\sin^22\phi}$, which is bigger than 1 unless
$\phi=0$. It is interesting to re-derive this result starting from
the spectral decomposition of $B_2$. Consider the Bell states:
$\ket{\Phi^{\pm}_z} = \frac{1}{\sqrt{2}}(\ket{00}\pm\ket{11})$,
$\ket{\Psi^{\pm}_z} = \frac{1}{\sqrt{2}}(\ket{01}\pm\ket{10})$.
$\ket{\psi(\phi)}$ is a linear combination of $\ket{\Phi^{+}_z}$
and $\ket{\Phi^{-}_z}$. If we take the unit vectors
$\underline{\mathbf{a}}$ in the $(x,y)$ plane, then
$\ket{\Phi^{+}_z}$ and $\ket{\Phi^{-}_z}$ must be associated to
opposite eigenvalues. But $\ket{\Phi^{+}_z}=\ket{\Phi^{+}_x}$ and
$\ket{\Phi^{+}_z}=\ket{\Psi^{+}_x}$: therefore, by taking the
unitary vectors in the $(y,z)$ plane, we can construct $B_2$ as
\[B_2\,=\,\lambda_1\,(P_{\Phi^+_x}-P_{\Phi^-_x})+
\lambda_2\, (P_{\Psi^+_x}- P_{\Psi^-_x})\,=\,
\lambda_1\,(P_{\Phi^+_z}-P_{\Psi^+_z})+ \lambda_2\,
(P_{\Phi^-_z}- P_{\Psi^-_z})\,.\] This way, the two vectors that
have a non-zero overlap with $\ket{\psi(\phi)}$ are associated to
the positive eigenvalues. The calculation of $S_2$ is not
difficult, using the fact that $\lambda_1^2+\lambda_2^2=2$ (see
Lemma \ref{l3} of Appendix A) and the standard maximization \ba
\max_\chi(A\cos\chi+B\sin\chi) &=&
\sqrt{A^2+B^2}\,.\label{maxcos}\ea We find indeed Horodecki's
value. Thus, to obtain this maximal violation of CHSH we took
advantage of the possibility of choosing the Bell states that are
orthogonal to $\ket{\psi(\phi)}$ as the states associated to the
negative eigenvalues of $B_2$. Now, this is precisely a
characteristic of two-qubit maximally entangled states that {\em
does not} generalize to three or more qubits. In fact, it is
well-known and easily verified that N-qubit GHZ states take the
form $\frac{1}{\sqrt{2}}(\ket{0^N}+\ket{1^N})$ only in one basis
(up to trivial relabelling). Therefore, for $N>2$, if we build
${\cal{B}}_N$ such that $\ket{\mbox{GHZ}_+}
=\frac{1}{\sqrt{2}}(\ket{0^N}+\ket{1^N})$ is associated to the
eigenvalue $\lambda$, then necessarily $\ket{\mbox{GHZ}_-}=
\frac{1}{\sqrt{2}} (\ket{0^N}-\ket{1^N})$ will be associated to
$-\lambda$.

\begin{figure}
\begin{center}
\epsfbox{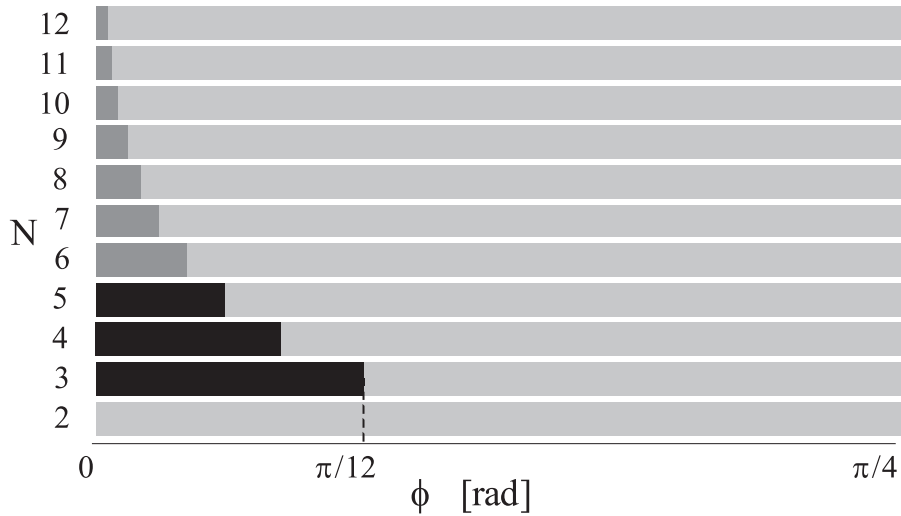} \caption{Ranges of $\phi$ for which
$\ket{\psi_N(\phi)}$ violates (brighter) and does not violate
(darker) the MK inequality. For $N>5$, the result is conjectured,
see (\ref{conj}).} \label{figphin}
\end{center}
\end{figure}

We now consider $S_N=\max_{\underline{a}} \moy{B_N(\underline{
\mathbf{a}})}_{\psi_N (\phi)}$, with $B_N$ a MK Bell operator,
for arbitrary $N\geq 3$. Using
$\sandwich{1^N}{B_N}{1^N}=(-1)^N\sandwich{0^N}{B_N}{0^N}$ we have
\ba\moy{B_N(\underline{\mathbf{a}})}_{\psi_N (\phi)}&=&
f_N(\phi)\,\sandwich{0^N}{B_N(\underline{\mathbf{a}})}{0^N}\,+\,
\sin 2\phi\,\mbox{Re}(\sandwich{1^N}{B_N(\underline{
\mathbf{a}})}{0^N}) \label{bna} \ea where
$f_N(\phi)=[\cos^2\phi+(-1)^N\sin^2\phi]$, that is 1 for $N$ even
and $\cos 2\phi$ for $N$ odd. The maximization of (\ref{bna})
over all possible choices of $\underline{\mathbf{a}}$ is not
evident for the following reason. We know that there are sets
$\underline{\mathbf{a}}$ that saturate the bound
$\mbox{Re}(\sandwich{1^N}{B_N}{0^N})= 2^{\frac{N-1}{2}}$; but for
these we find $\sandwich{0^N}{B_N}{0^N}= 0$. Similarly, the sets
$\underline{\mathbf{a}}$ that saturate the bound
$\sandwich{0^N}{B_N}{0^N}= 1$ give
$\mbox{Re}(\sandwich{1^N}{B_N}{0^N})=0$. Let us try to guess the
maximum of (\ref{bna}) using the insight provided by the spectral
decomposition of $B_N$ discussed in section 2 above. A natural
first guess would be $B_N=2^{\frac{N-1}{2}} \big(P_{\mbox{GHZ}_+}-
P_{\mbox{GHZ}_-}\big)$, that is
$\mbox{Re}(\sandwich{1^N}{B_N}{0^N}) =2^{\frac{N-1}{2}}$. This
choice gives $S_N^{(g)}(\phi)=2^{\frac{N-1}{2}}\sin 2\phi$.
Numerical evidence suggests that this is indeed the maximum of
(\ref{bna}) whenever $S_N(\phi)\geq 1$. This provides a criterion
for violation of MK inequalities: \ba S_N(\phi)\,>\, 1 \;\;
\Longleftrightarrow\;\; \sin 2\phi \,>\, 2^{-\frac{N-1}{2}}&&
\left(\begin{array}{ll}
N=3,4,5\,:& \mbox{numerically verified} \\
N>5\,:&\mbox{conjectured}\end{array} \right)\,. \label{conj} \ea
Thus there exist {\em pure} entangled states that do not violate
the MK inequality. Let's define $\phi_{N}$ as the value of $\phi$
at which $\ket{\psi_N(\phi)}$ ceases to violate the MK inequality:
we have $\phi_2=0$, and $\sin 2\phi_{N} =2^{-\frac{N-1}{2}}$ for
$N\geq 3$, within the validity of (\ref{conj}). There is a clear
discontinuity in the behaviour of $\phi_N$ between $N=2$ and
$N=3$, as illustrated in figure \ref{figphin}: $\phi_N$ jumps from
0 for $N=2$ to $\frac{\pi}{12}$ for $N=3$, then starts decreasing
again for higher $N$. This analysis suggests that the MK
inequalities, and more generally the family of Bell's
inequalities with two observables per qubit, may not be the
"natural" generalization of the CHSH inequality to more than two
qubits. Whether a more suitable inequality exists is an open
question.

To conclude this section, let us briefly come back to the problem
of maximizing (\ref{bna}). The guess $S_N^{(g)}(\phi)=
2^{\frac{N-1}{2}}\sin 2\phi$ cannot be correct for all $\phi$: in
fact, in the limit of small $\phi$'s, $\ket{\psi_N(\phi)}$
approaches $\ket{0^N}$ and therefore $S(\phi)$ should converge to
1. To avoid this problem, we modify slightly our guess to have \ba
S_N^{(g)}(\phi)&=& \max\left[ 2^{\frac{N-1}{2}}\sin 2\phi\,,\,
f_{N}(\phi) \right]\,. \label{sguess}\ea Obviously $S_N\geq
S_N^{(g)}$, since we know that $S_N^{(g)}$ can be reached.
Numerical estimates for $N=3,4,5$ prove that this guess is
extremely good. Actually, for $N=4,5$ we found $S_N(\phi)=
S_N^{(g)}(\phi)$ for all $\phi$ to within the accuracy of the
calculation; for $N=3$, the same holds for all $\phi$ but a small
range of values around $\bar{\phi}$ defined by $2\sin
2\bar{\phi}=\cos 2\bar{\phi}$ (the maximal difference is
$S_3-S_3^{(g)}\approx 0.002$).

\subsection{Bounds on the violation of Mermin's inequality}

As a second application, we show how the knowledge of the spectral
decomposition of Bell operators provides bounds to estimate the
violation of Mermin's inequality for any three-qubit state $\rho$.
Some results are similar to those found independently by Zukowski
and Brukner \cite{zb}.

We consider a three-qubit Bell operator
${\cal{B}}_3(\underline{\mathbf{a}})$. According to Theorem
\ref{propeigen} one can always find a basis such that its eight
eigenstates are $\ket{\Psi_{1,8}}=\ket{\Psi_{000}(0,\pi)}=
\frac{1}{\sqrt{2}}(\ket{000}\pm\ket{111})$,
$\ket{\Psi_{2,7}}=\ket{\Psi_{001}(0,\pi)}=
\frac{1}{\sqrt{2}}(\ket{001}\pm\ket{110})$,
$\ket{\Psi_{3,6}}=\ket{\Psi_{010}(0,\pi)}=
\frac{1}{\sqrt{2}}(\ket{010}\pm\ket{101})$,
$\ket{\Psi_{4,5}}=\ket{\Psi_{011}(0,\pi)}=
\frac{1}{\sqrt{2}}(\ket{011}\pm\ket{100})$. Note that the four
angles $\theta_{\Omega}$ that are unconstrained can be chosen to
be 0 without loss of generality, since this amounts to a
redefinition of the global phases of $\ket{0}_A$, $\ket{0}_B$
etc. Consequently in this basis \ban {\cal{B}}_3 &=&
\lambda_1\,(P_1-P_8)\,+\,\lambda_2\,(P_2-P_7)\,+
\lambda_3\,(P_3-P_6)\,+ \lambda_4\,(P_4-P_5)\,=\\ &=&
\mu_{++++}\si_{xxx}+\mu_{-++-}\si_{xyy}+\mu_{-+-+}\si_{yxy}
+\mu_{--++}\si_{yyx}\ean with $\si_{xxx}=\si_x\otimes
\si_x\otimes\si_x$ etc., and with $\mu_{s_1s_2s_3s_4}=\frac{1}{4}
(s_1\lambda_1+s_2\lambda_2+s_3\lambda_3+s_4\lambda_4)$. For a
given three-qubit state $\rho$ \ba \mbox{Tr}({\cal{B}}_3\,\rho)&=&
\mu_{++++}t_{xxx}+\mu_{-++-}t_{xyy}+\mu_{-+-+}t_{yxy}
+\mu_{--++}t_{yyx}\ea with the standard notation $t_{xxx} =
\mbox{Tr}(\rho\,\si_{xxx})$ etc. Our final purpose is to estimate
$S_{\rho}=\max_{\underline{a}} \mbox{Tr}({\cal{B}}_3\,\rho)$ for
any $\rho$. If $\rho$ is given, one must find both the good
eigenvectors and the good eigenvalues of ${\cal{B}}_3$. The
optimization of the eigenvalues is performed by varying the
parameters $\mu$; we discuss it in the next paragraph for the
Mermin's operator $B_3$. To optimize the eigenvectors means to
define the axes $x$ and $y$ for each qubit. Note that when the
basis of eigenvectors is optimized only four number $t_{ijk}$ will
come into play, thus sharpening the condition obtained by
Zukowski and Brukner \cite{zb} that involved eight of these
numbers.

While the system of eigenvectors is the same for all Bell
operators of the form (\ref{polynomial}), the eigenvalues and
their properties obviously depend on the operator that is
considered. We restrict our discussion to the Mermin operator
$B_3$ given by (\ref{recurrence}). It can then be shown that the
eigenvalues must satisfy \ba\mbox{Tr}(B_3^2)&=&
8\label{condtrace}\ea (see Lemma \ref{l3} in Appendix A). This
leads to $\mu_{++++}^2+\mu_{-++-}^2 +\mu_{-+-+}^2
+\mu_{--++}^2=1$. Therefore, we can let
$\mu_{++++}=\cos\alpha\cos\beta$,
$\mu_{-++-}=\cos\alpha\sin\beta$,
$\mu_{-+-+}=\sin\alpha\cos\gamma$ and
$\mu_{--++}=\sin\alpha\cos\gamma$, and maximize over $\alpha$,
$\beta$ and $\gamma$. By using thrice the maximization
(\ref{maxcos}) we find \ba S_{\rho}&\leq&S^{+}_{\rho}\,=\,
\max_{\{x,y\}}\,\sqrt{t^2_{xxx}+t^2_{xyy}+t^2_{yxy}+t^2_{yyx}}\,.
\label{upper} \ea This bound would in fact be exact if there were
no constraint on the eigenvalues other than (\ref{condtrace}).
However, starting from the eigenvalues as they are given in
statement 2 of Theorem \ref{propeigen}, one finds by inspection
that the eigenvalues are bound to fulfill some other conditions,
like $\frac{(\lambda_3^2+\lambda_1^2-2)(\lambda_3^2+
\lambda_4^2-2)}{(\lambda_3^2+\lambda_2^2-2)}\,=\,
2\sin^2(\alpha_1-\alpha_1')$. To see that such a condition is
indeed an additional constraint, choose $\lambda_2=\lambda_3=1$,
which is of course a possible choice. The rhs is bounded by 2,
thus for the lhs not to diverge we must also have (say)
$\lambda_4=1$; but then the condition (\ref{condtrace}) forces
$\lambda_1=1$ too. In conclusion, if two eigenvalues are equal to
1, all the eigenvalues must be equal to 1.

Since such constraints are not easy to handle, it is interesting
to provide a lower bound on $S_{\rho}$. A non-trivial one is
obtained by simply choosing one possible realization of the
eigenvalues, namely $\lambda_1=\pm 2$, which due to
(\ref{condtrace}) implies $\lambda_2=\lambda_3=\lambda_4=0$. This
gives \ba S_{\rho}&\geq&S^{-}_{\rho}\,=\,
\demi\,\max_{\{x,y\}}\,\left|t_{xxx}-t_{xyy}-t_{yxy}-t_{yyx}\right|\,.
\label{lower} \ea In most cases, we still have to rely on a
computer program to calculate bounds (\ref{upper}) and
(\ref{lower}). For these bounds, the optimization bears on nine
parameters (for each qubit, two parameters define the $(x,y)$
plane and a third one fixes the axes in the plane); while a
direct optimization of $\moy{B_3}$ bears on twelve parameters
(two unit vectors per qubit).

Let's conclude by a discussion of the quality of the bounds
(\ref{upper}) and (\ref{lower}), based on some examples. Consider
first the family of states $\ket{\psi_3(\phi)}$ (\ref{psinphi}).
From $\phi=\frac{\pi}{4}$ down to $\phi\approx\frac{\pi}{10}$ we
find $S^{-}=S=S^{+}\,=\,2\sin 2\phi$. For smaller values of
$\phi$: (i) $S^{-}\,=\,\max(2\sin 2\phi,\cos 2\phi)$; that is,
$S^{-}$ corresponds to $S_3^{(g)}$ defined in (\ref{sguess}).
(ii) As we discussed above, the exact value $S$ essentially
follows $S^{-}$, but for small deviations. (iii) The upper bound
$S^{+}$ increases again, from $S^{+}(\frac{\pi}{10})\approx
1.175$ up to $S^{+}(0)=\sqrt{2}$. Thus this bound turns out to be
too rough in the region $\phi\leq\frac{\pi}{12}$, where
$\ket{\psi_3(\phi)}$ ceases to violate Mermin's inequality.

In table \ref{tables3} we give $S_{\rho}$ and the bounds
$S^+_\rho$ and $S^-_\rho$ for some other states. All the possible
cases $S^{-}=S<S^{+}$, $S^{-}<S=S^{+}$ and $S^{-}<S<S^{+}$ are
present. On all these examples, we see that at least one bound is
very close, if not identical, to the exact value.

\begin{table}
  \centering
\begin{tabular}{|c|c|c|c|}
 \hline State & $S^-_\rho$ & $S_{\rho}$ & $S^+_\rho$ \\ \hline
    $\ket{W}=\ket{011}+\ket{101}+\ket{110}$ & 1.516 & 1.523 & 1.527 \\
    $\cos\frac{\pi}{5}\ket{000}+\sin\frac{\pi}{5}\ket{W}$ & 1.669 & 1.669 & 1.68 \\
    $\cos^2\frac{\pi}{5}\ket{000}+\cos\frac{\pi}{5}\ket{001}+\sin\frac{\pi}{5}\ket{111}$ & 1.425 & 1.431 & 1.431 \\
    $\ket{0}(\ket{00}+\ket{11})$ & 1 & $\sqrt{2}$ & $\sqrt{2}$ \\ \hline
  \end{tabular}
  \caption{$S_{\rho}$ and the bounds $S^+_\rho$ and $S^-_\rho$ for some three-qubit states
  (states not-normalized).}\label{tables3}
\end{table}

\section{Conclusion}

Bell's inequalities for systems of more than two qubits are the
object of renewed interest, motivated by the fact that
entanglement between more than two quantum systems is becoming
experimentally feasible. A link between Bell's inequalities and
the security of quantum communication protocols has also been
stressed recently \cite{sca}.

Here we focused on inequalities obtained by measuring two
observables per qubit, and we gave the spectral decomposition of
the corresponding operators. With this tools, we studied the
violation of Mermin-Klyshko inequalities for some states that
exhibit N-qubit entanglement. We proved numerically for $N=3,4,5$,
and we conjectured for all $N$, that there exist pure entangled
states that do not violate these inequalities.

We acknowledge financial support from the Swiss FNRS and the
Swiss OFES within the European project EQUIP (IST-1999-11053).

\section*{Appendix A: Relationships between $B_n$ and $B_n'$}

We derive in this Appendix some properties of the MK operators
$B_n$ and $B_n'$ that were not discussed in previous publications
\cite{belin,helle}. Lemma \ref{l1} was demonstrated independently,
and with different mathematical tools, in \cite{wer}.

\newtheorem{l1}[l4]{Lemma}
\begin{l1}
$B_n^2={B_n'}^2$ for all $n$. \label{l1}
\end{l1}
{\em Proof:} From (\ref{recurrence}) we have \ba B_n^2&=&
\demi\,(1+\mathbf{a}_n\cdot\mathbf{a}_n')\one\otimes{B}^2_{n-1}
\,+\, \demi\,
(1-\mathbf{a}_n\cdot\mathbf{a}_n')\one\otimes{B'}^2_{n-1}
\,-\,\frac{i}{2}\,\si_{a_n\wedge
a_n'}\otimes\com{B_{n-1}}{B_{n-1}'}\,.\label{bncarreprov}\ea
${B_n'}^2$ is obtained by exchanging the primed with the
non-primed objects. Therefore \ban
B_n^2-{B_n'}^2&=&\mathbf{a}_n\cdot\mathbf{a}_n'\,
\big(B_{n-1}^2-{B'}^2_{n-1}\big)\,\propto\,
\big(B_{2}^2-{B_{2}'}^2\big)\,=\,0\ean since it can be calculated
explicitly that $B_{2}^2={B_{2}'}^2=\one+\si_{a_2\wedge
a_2'}\otimes\si_{a_1\wedge a_1'}$. $\diamond$

\newtheorem{l3}[l4]{Lemma}
\begin{l3}
The explicit expressions for $B_n^2$ and for the commutator
$\com{B_n}{B_n'}$ are given respectively by (\ref{bncarre}) and
(\ref{commutator}). The anticommutator is
$\left\{B_n,B_n'\right\}= 2(\mathbf{a}_n\cdot\mathbf{a}_n')
...(\mathbf{a}_1\cdot\mathbf{a}_1')\,\one$. As a corollary, note
that $\mbox{Tr}(B_n^2)= 2^n$. \label{l3}
\end{l3}

{\em Proof:} Lemma \ref{l1} allows us to rewrite
(\ref{bncarreprov}) as \ba B_n^2&=& \one\otimes{B}^2_{n-1}
\,-\,\frac{i}{2}\,\si_{a_n\wedge
a_n'}\otimes\com{B_{n-1}}{B_{n-1}'}\,.\ea Another standard
calculation from (\ref{recurrence}) leads to \ba
\com{B_n}{B_n'}&=& \one\otimes\com{B_{n-1}}{B_{n-1}'}\,+\,
2i\si_{a_n\wedge a_n'}\otimes{B}^2_{n-1}\,.\ea The structure of
these two equations can be best seen by introducing the notations
$B_n^2\equiv P_n$, $\com{B_n}{B_n'}\equiv 2iQ_n$ and
$\si_{a_n\wedge a_n'}\equiv\Sigma_n$. We have then
\[
\left\{\begin{array}{lcl} P_n&=&\one\otimes P_{n-1}+
\Sigma_n\otimes Q_{n-1}\\
Q_n&=&\one\otimes Q_{n-1}+ \Sigma_n\otimes P_{n-1}
\end{array}\right.\,.
\]
The recursive solution is a matter of patience. Using
$P_2=\one+\Sigma_2\otimes\Sigma_1$ and
$Q_2=\one\otimes\Sigma_1+\Sigma_2\otimes\one$ we find \ba
B_n^2={B_n'}^2&=&\one_{2^n}\,+\, \sum_{i<j}\si_{a_i\wedge
a_i'}\si_{a_j\wedge a_j'}\,+\,\sum_{i<j<k<l}\si_{a_i\wedge
a_i'}\si_{a_j\wedge
a_j'}\si_{a_k\wedge a_k'}\si_{a_l\wedge a_l'} \,+\,... \label{bncarre}\\
\com{B_n}{B_n'}&=&2i\,\left(\sum_{i}\si_{a_i\wedge
a_i'}\,+\,\sum_{i<j<k}\si_{a_i\wedge a_i'}\si_{a_j\wedge
a_j'}\si_{a_k\wedge a_k'} \,+\,...\right)\label{commutator}\ea
where the dots indicate the sums over all products of an even,
resp. an odd, number of $\si_{a_i\wedge a_i'}$.

Finally, the anticommutator is also found through a direct
calculation from (\ref{recurrence}): \ban
\left\{B_n,B_n'\right\}&=&\frac{1}{4}
\Big(2(\one+\left\{\si_{a_n},\si_{a_n'}\right\}\otimes
\left\{B_{n-1},B_{n-1}'\right\}) -
2(\one-\left\{\si_{a_n},\si_{a_n'}\right\}\otimes
\left\{B_{n-1},B_{n-1}'\right\})\,+\\
&&+\,\left\{\si_{a_n}+\si_{a_n'},\si_{a_n}-\si_{a_n'}\right\}\otimes
\underbrace{\left({B_{n-1}'}^2-B_{n-1}^2\right)}_{=0} \Big)\,=\\
&=& \demi\left\{\si_{a_n},\si_{a_n'}\right\}\otimes
\left\{B_{n-1},B_{n-1}'\right\}\,=\,
(\mathbf{a}_n\cdot\mathbf{a}_n')\one\otimes
\left\{B_{n-1},B_{n-1}'\right\}\,.\ean The conclusion follows from
$\left\{B_2,B_2'\right\}=
2(\mathbf{a}_2\cdot\mathbf{a}_2')(\mathbf{a}_1\cdot\mathbf{a}_1')\,\one_4$.
$\diamond$

\section*{Appendix B: Some results about the maximal eigenvalue}

In the main text we exhibited a set of eigenvectors of $B_n$ that
have a remarkable symmetry. But this set would lose much of its
interest if all eigenvalues were degenerate. Here we show that at
least in one case (which is an interesting one) we can be sure
that there are non-degenerate eigenvalues.

Let's first sort the eigenvalues of $B_n$ in decreasing order:
$\lambda_1\geq...\geq\lambda_{2^n}$. By virtue of lemma \ref{l4},
$\lambda_{k}=-\lambda_{2^n-k+1}$. In particular,
$\mbox{Tr}(B_n^2)=2(\lambda_1^2+\lambda_2^2+...+\lambda_{2^{n-1}}^2)$.
On the other side, we have noticed in lemma \ref{l3} that
$\mbox{Tr}(B_n^2)= 2^n$, whence \be
\lambda_1^2+\lambda_2^2+...+\lambda_{2^{n-1}}^2=
2^{n-1}\,.\label{lambdas0}\ee In particular, if
$\lambda_1=2^{\frac{n-1}{2}}$, then all the other eigenvalues
(except of course $-\lambda_1$) are zero. In general,
$\lambda_1\geq 1$, where the equality holds only in the
"classical" case $\lambda_1=...=\lambda_{2^{n-1}}=
-\lambda_{2^{n-1}+1}=...=-\lambda_{2^{n}}=1$. Equality
(\ref{lambdas0}) implies \be
\lambda_1^2+\lambda_2^2\,\leq\,2^{n-1},\label{lambdas}\ee whence
the following:
\newtheorem{l5}[l4]{Lemma}
\begin{l5}
let $\lambda_1$ and $\lambda_2$ be the two greatest eigenvalues
of $B_n$. If $\lambda_1>2^{\frac{n}{2}-1}$, then it is
non-degenerate, and moreover
$\lambda_2<2^{\frac{n}{2}-1}$.\label{l5}
\end{l5}
Now, if $\rho$ is a n-qubit state exhibiting m-party
entanglement, $m\leq n$, it can be shown that
$\moy{B_n}_{\rho}\leq 2^{(m-1)/2}$ \cite{helle,wer}. Thus
$\lambda_1>2^{\frac{n}{2}-1}$ for $B_n$ means that we have a
n-qubit violation. So our last lemma reads: if the parameters
$\underline{\mathbf{a}}$ of $B_n$ are such that a n-qubit
violation is possible, the maximal eigenvalue of $B_n$ is
non-degenerate. Actually we can prove even more:
\newtheorem{l6}[l4]{Lemma}
\begin{l6}
If $\lambda_1>2^{\frac{n}{2}-1}$, one cannot find two orthogonal
states that both satisfy the condition
\be\moy{B_n}_{\psi}>2^{\frac{n}{2}-1}\,.\label{cond}\ee Due to
lemma \ref{l4}, the same holds for the condition
$\moy{B_n}_{\psi}<-2^{\frac{n}{2}-1}$. \label{l6}
\end{l6}
To prove this lemma, we determine a necessary conditions for a
state  $\ket{\psi}$ to satisfy (\ref{cond}). Let's decompose
$\ket{\psi}$ on the basis of the eigenvectors of $B_n$:
$\ket{\psi}\,=\,\sum_{k=1}^{2^n}\sqrt{p_i}\,\ket{\Psi_i}$ where
$\ket{\Psi_i}$ is an eigenvector of $B_n$ for the eigenvalue
$\lambda_i$. With these notations \ban
\sandwich{\psi}{B_n}{\psi}\,=\,
\sum_{i=1}^{2^n}p_i\,\lambda_i&\leq&p_1\lambda_1+(1-p_1)\lambda_2\,.\ean
Therefore if $p_1\lambda_1+(1-p_1)\lambda_2\leq 2^{n/2-1}$, the
requirement (\ref{cond}) cannot be satisfied. In other terms, a
necessary condition for (\ref{cond}) to be satisfied is \ba
p_1\,>\,
\frac{2^{\frac{n}{2}-1}-\lambda_2}{\lambda_1-\lambda_2}\,=\,
\frac{1-\mu_2}{\mu_1-\mu_2}\equiv\,\bar{p}\ea (we introduced the
notation $2^{n/2-1}\mu_i=\lambda_i$ in order to show that
$\bar{p}$ does not depend explicitly on the number of qubits
$n$). It can be shown using (\ref{lambdas}) that $\frac{1}{2}<
\bar{p}\leq 1$. The limiting case $\bar{p}=1$ corresponds to
$\lambda_1=2^{n/2-1}$, in which case of course (\ref{cond}) cannot
be satisfied. The roughest criterion that we can state is
therefore the following: given $B_n$ such that
$\lambda_1>2^{n/2-1}$, a state $\ket{\psi}$ cannot satisfy
(\ref{cond}) if $p_1=\left|\braket{\Psi_1}{\psi}
\right|^2\leq\frac{1}{2}$. This criterion is enough to conclude
the proof of lemma \ref{l6}.


\begin{thebibliography}{10}
\bibitem{note0} Usually, the CHSH operator is written without the
factor $\demi$ in front of it, so that the Bell's inequality is
$S\leq 2$.
\end{thebibliography}
\end{document}